\documentclass[aps, prb, preprint]{revtex4-1}

\usepackage{graphicx,color}
\usepackage{ulem}

\usepackage{dcolumn}
\usepackage{bm}
\usepackage{amsmath}
\usepackage{ulem}

\begin{document}





\title{Nanofaceting as a stamp for periodic graphene charge carrier modulations}


\author{M. Vondr\'a\v{c}ek}  
\affiliation{Institute of Physics, Academy of Sciences of the
Czech Republic, CZ-182 21 Praha 8, Czech Republic}

\author{D. Kalita}
\affiliation{D\'epartement Nanosciences, CNRS, Institut N\'eel, F-38042 Grenoble, France}

\author{M. Ku\v{c}era}
\affiliation{Institute of Physics, Academy of Sciences of the
Czech Republic, CZ-182 21 Praha 8, Czech Republic}

\author{L. Fekete}
\affiliation{Institute of Physics, Academy of Sciences of the
Czech Republic, CZ-182 21 Praha 8, Czech Republic}

\author{J. Kope\v{c}ek}
\affiliation{Institute of Physics, Academy of Sciences of the
Czech Republic, CZ-182 21 Praha 8, Czech Republic}

\author{J. Lan\v{c}ok}
\affiliation{Institute of Physics, Academy of Sciences of the
 Czech Republic, CZ-182 21 Praha 8, Czech Republic}

\author{J. Coraux}
\affiliation{D\'epartement Nanosciences, CNRS, Institut N\'eel, F-38042 Grenoble, France}

\author{V. Bouchiat}
\affiliation{D\'epartement Nanosciences, CNRS, Institut N\'eel, F-38042 Grenoble, France}

\author{J. Honolka}
\affiliation{Institute of Physics, Academy of Sciences of the
Czech Republic, CZ-182 21 Praha 8, Czech Republic}
\email{honolka@fzu.cz}

\date{\today}

\begin{abstract}
The exceptional electronic properties of monoatomic thin graphene sheets triggered numerous original transport concepts, pushing quantum physics into the realm of device technology for electronics, optoelectronics and thermoelectrics. At the conceptual pivot point is the particular two-dimensional massless Dirac fermion character of graphene charge carriers and its volitional modification by intrinsic or extrinsic means. Here, interfaces between different electronic and structural graphene modifications promise exciting physics and functionality, in particular when fabricated with atomic precision.\\
In this study we show that quasiperiodic modulations of doping levels can be imprinted down to the nanoscale in monolayer graphene sheets. Vicinal copper surfaces allow to alternate graphene carrier densities by several $10^{13}$ carriers per cm$^{2}$ along a specific copper high-symmetry direction. The process is triggered by a self-assembled copper faceting process during high-temperature graphene chemical vapor deposition, which defines interfaces between different graphene doping levels at the atomic level.
\end{abstract}


\maketitle
Graphene, a simple two-dimensional honeycomb arrangement of $sp^2$-hybridised carbon atoms, is hailed for its exceptional electronic environment, forcing charge carriers to propagate analogous to relativistic massless particles~\cite{Castro-Neto-Overview}. Its potential to revolutionize standard silicon-based electronics is widely recognised, provided that material properties like local defects, honeycomb rotational order or electronic doping can be controlled and engineered at hand down to the nanometer scale, i.e. at or beyond the limits of standard top-down state-of-the-art nanofabrication techniques.\\
\indent \indent Immense progress was achieved in recent years on fabricating high-quality homogeneous graphene sheets with small defect densities, reaching high carrier mobilities up to several 100.000 cm$^{2}$/Vs. However, the crucial step towards a targeted realisation of heterogeneous graphene properties, mostly relying on lithography techniques, systematically faces spurious degradation of the structure and performance of devices. Yet, heterogeneous properties majorly widen the options for electronics and for experiments on exciting fundamental physics: 1D grain boundaries between different honeycomb lattice orientations can be exploited to achieve variable bandgaps for optoelectronics in otherwise semi-metallic graphene~\cite{Yazyev}, to tune carrier mobilities~\cite{Tuan2013}, or to introduce spin degrees of freedom~\cite{Cervenka2009}. Local control over graphene electronic doping is of particular interest, since it allows to induce $p-n$ junctions as a basis for transistor functionality~\cite{Chiu, Williams}. Moreover, when reduced to a small scale, such junctions should bring to life very fundamental prospects of relativistic quantum mechanics such as the so-called Dirac-fermion optics~\cite{Allain2011}, where refraction of electron and hole waves at $p-n$ transitions is governed by doping levels and their spatial abruptness~\cite{Cheianov2007, Cayssol2009, Park2008, Young2009}. A hallmark in this field is the predicted Klein tunneling effect~\cite{Katsnelson2006}.\\
\indent Supporting metallic surfaces are rich playgrounds for these concepts, moreover offering the prospect of large scale production of high-quality graphene via chemical vapor deposition (CVD). Indeed, metals may exhibit coexisting surface terminations with different interaction potentials and the potential to trigger variations in graphene doping~\cite{Walter}. They allow the formation of graphene with different crystallographic orientations~\cite{Huang2011}, different kinds of grain boundaries between domains, and domains with various doping levels~\cite{Starodub2011,  Avila2013}.\\
\indent In this article we report an unprecedented 1D quasiperiodic modulation of graphene electron doping, probed by spatial mapping of the electronic band structure in wave-vector-resolved photoemission microscopy ($k$-PEEM).
Sampling local topography and diffraction, we show that a nanometer-scale periodic structuration and electronic doping of several $10^{13}$ carriers per cm$^{2}$ can be achieved straightforwardly in graphene, as-grown by CVD on high-index vicinal copper. The pattern consists of a roof-top-like alternation of Cu facets of distinctive symmetries, formed by surface energy minimization at the atomic scale, which drives copper and carbon mass-transfers during high-temperature CVD. The general concept of this work, which avoids any lithography processing steps, can be extended towards other chemical vapor deposited 2D systems of current interest such as semiconducting transition metal dichalcogenides, e.g. MoS$_2$, insulating hexagonal boron nitride (h-BN) monolayers, and respective hybrid structures.\\

{\noindent \bf \large Results}\\
{\noindent \bf Graphene sample fabrication and vicinal copper foil characterisation.}

\begin{figure}
\begin{flushleft}
\center \includegraphics[width=100mm]{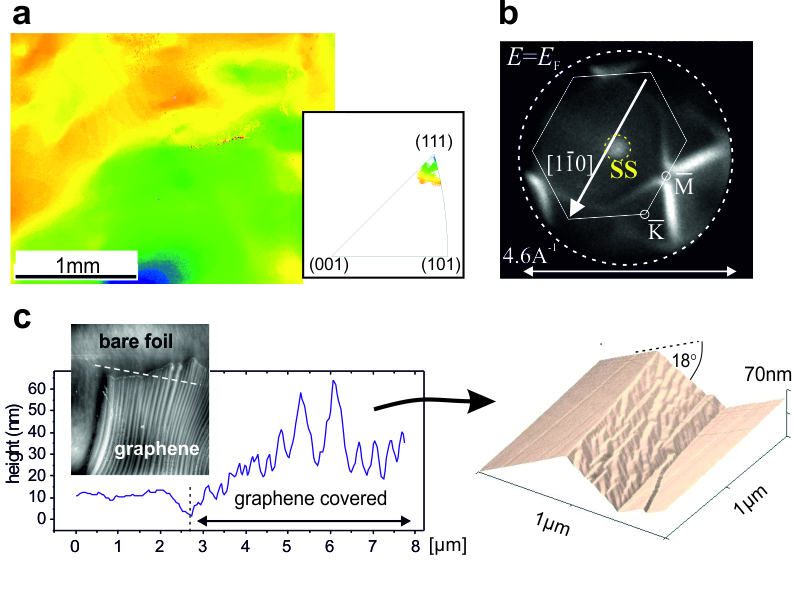}
\caption{{\bf Structure and morphology of the supporting Cu-foils.}
{\bf (a)} EBSD map of an area about $2\,$mm$\times3\,$mm on the Cu-foil. The colors correspond to orientations plotted in the inverse pole figure in the inset.
{\bf (b)} $k$-PEEM image of the foil measured at the Fermi level $E_{\text{F}}$ after graphene removal. The $k$-space pattern reveals a tilted hexagonal Cu(111) surface Brillouin zone according to a homogeneous vicinal character of the Cu foil, inclined by the rotation vector [$1\overline{1}0$]. In the center the Mahan cone of the Cu(111) surface state (SS) is visible.  
{\bf (c)} AFM images and height profile over the edge of a graphene island. In the 3D plot a repetitive $\mu$m scale roof-top structure emerges with canting angles of $(18\pm4)^{\circ}$ and a onesided complex facet structure.
}
\label{fign0}
\end{flushleft}
\end{figure}

\noindent Single-layer graphene was prepared on commercial Cu foils at growth temperatures of $1020\,^{\circ}$C, following a pulsed CVD method, which prevents the formation of multilayer patches at the nucleation centers as described in an earlier work~\cite{Han}. In continuous CVD, 2$^{\text{nd}}$ and 3$^{\text{rd}}$ layer patches are known to grow from below due to carbon atoms dissolved in bulk copper.\\
Prior to surface sensitive photoemission electron microscopy (PEEM), low-energy electron diffraction (LEED), and x-ray photoemission spectroscopy (XPS) measurements under ultra-high vacuum (UHV) conditions, the samples were annealed {\it in-situ} at temperatures of $400\,^{\circ}$C. \\
Electron backscattering diffraction (EBSD) in Fig.~\ref{fign0}a reveals a Cu foil crystal orientation close to (111), however with local variations in the orientation defined by the color coding. 
On the millimeter scale these variations correspond to angles smaller than $\pm3^{\circ}$, due to the waviness of the Cu foil. 
The inclination of the (111) direction with respect to the surface normal is directly visible in the $k$-PEEM pattern in Fig.~\ref{fign0}b, showing a high-index vicinal (111) cut of the copper`s Fermi surface as developed after graphene removal by a mild Ar$^+$ ion sputtering followed by $300\,^{\circ}$C annealing in UHV. 
The Fermi surface cut is tilted towards the labeled $\overline{\text{M}}$ point according to the rotation vector [$1\overline{1}0$] indicated in the figure. Typical for photon excitation energies $h\nu=21.2\,$eV, the Mahan cone of the Cu(111) surface state (SS) is detected, which is shifted away from the center of the (111) orientation against the tilting direction~\cite{Lobo, Baumberger}.\\

{\noindent \bf Faceting process}. Pulsed CVD leads to a characteristic graphene island morphology described in Fig.~\ref{fign0}c. Atomic force microscopy (AFM) on larger graphene islands shows a characteristic stripe structure due to a pronounced roof-top shaped height modulation with varying canting angles of $(18\pm4)^{\circ}$. Looking more closely, the roof-top structure is asymmetric and reveals a one sided complex faceted substructure on the nanometer scale. On the bare copper foil the roof-top modulation is absent, suggesting a graphene growth induced restructuring process, potentially related to the recently proposed feedback mechanism between the growing graphene and underlying mobile Cu atoms~\cite{Wilson, Hayashi}. \\

\begin{figure}
\begin{flushleft}
\center \includegraphics[width=110mm]{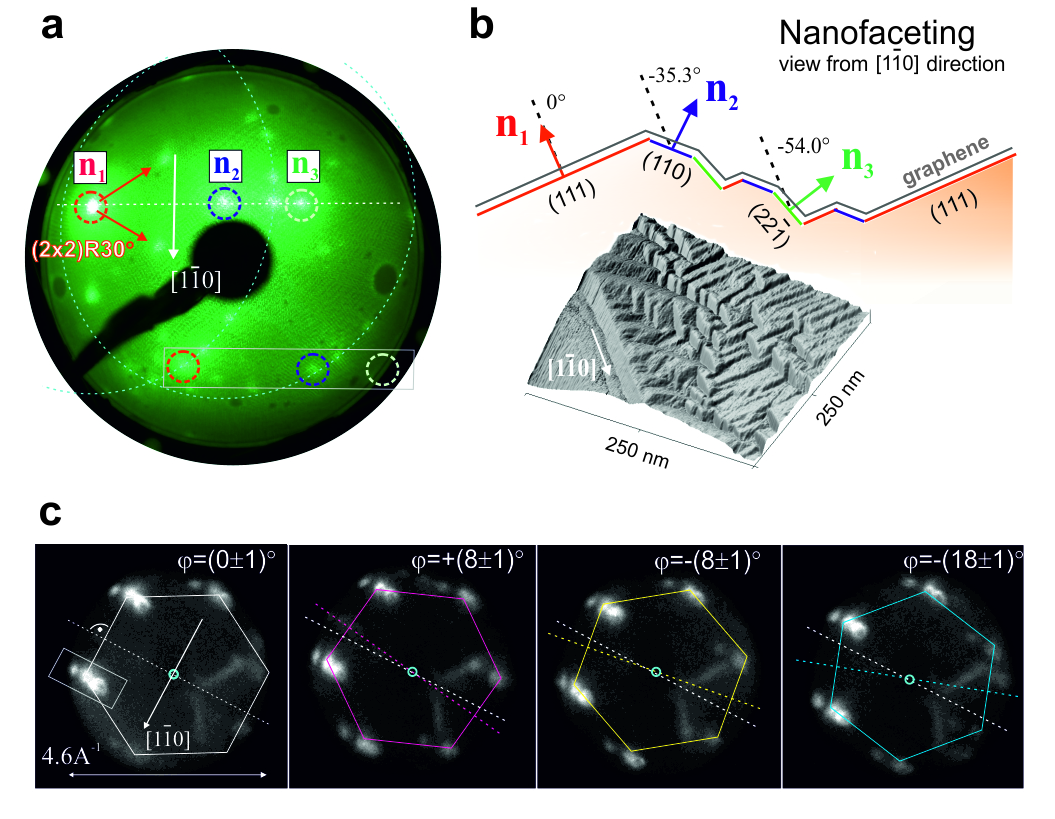}
\caption{{\bf Graphene orientation and geometry of nanofacets.}
{\bf (a)} The LEED pattern exhibits three specular reflections $n_{1}$, $n_{2}$, and $n_{3}$ aligned perpendicular to the [$1\overline{1}0$] direction, which correspond to crystal facets (111), (110), and ($22\overline{1}$). $n_{1}$ is defined by an oxygen p$(2\times2)$ superlattice, $30^{\circ}$ rotated with respect to the underlying Cu(111) facet (see red unit cell vectors in the LEED image).
Hexagonal LEED spots of rotational graphene domains translate to all three facets as concentrical circles, obeying the geometry of specular reflections (see example of a triplet graphene spot in the box). 
{\bf (b)} Faceting geometry viewed against the [$1\overline{1}0$] direction according to the STM image below.
{\bf (c)} Left to right: $k$-PEEM images at different positions on a large graphene patch, showing rotational domains defined by the angle $\varphi$.  
Independent of rotation angles $\varphi = 0^{\circ}$, $\pm 8^{\circ}$ and $+18^{\circ}$, graphene hexagons always show triplet replicas along the direction perpendicular to [$1\overline{1}0$] (see example of a triplet graphene spot in the box). 
}
\label{fign1}
\end{flushleft}
\end{figure}

Reciprocal space methods LEED and $k$-PEEM reveal further information on the graphene morphology and its domain orientations. Analysing energy-dependent LEED patterns shown in Supplementary Fig. S1, we can identify three specular spots. Specular spots correspond to elastically backscattered electron beams from local planes (zero order scattering of incoming beam), and thus for a uniformly flat surface one would expect only one. Here, they define three distinct facets with local surface normals $n_{1}$, $n_{2}$, and $n_{3}$ aligned perpendicular to the [$1\overline{1}0$] direction. $n_{1}$ has by far the largest intensity and thus dominates the surface area. Position-dependent LEED showed that the local surface normals are homogeneously oriented over millimeter scales on the copper foil. The diffraction LEED signal (first order scattering) in Fig.~\ref{fign1}a, which averages over a $1.5\,$mm spot on the sample, shows a few graphene rotational domains at once. Hexagonal LEED patterns of differently oriented coplanar graphene domains are expected to lay on a concentric circle around the supporting surface normal $n$~\cite{Walter, Hass}. In our case, each of the three facets reproduce the same rotational domain hexagons on respective concentric circles, generating a characteristic triplet of replica spots (white box, showing one example domain spot on the three surface normals). For reasons of clarity in Fig.~\ref{fign1}a we only indicate two of those circles corresponding to $n_{1}$ and $n_{2}$. \\
The facets are resolved in detail by STM in Fig.~\ref{fign1}b which unveils a length scale and shape reminiscent of CVD grown graphene on vicinal Ir(332)~\cite{Srut} or polycrystalline copper~\cite{Rasool}. Faceted surfaces typically self-assemble under the influence of monolayer coverages of adsorbates such as oxygen, sulfur, or metals as a result of anisotropies in the surface free energy [see e.g. Ref.~\cite{Ertl} for an overview]. In our case both carbon and oxygen seem to play a role, since local XPS on large graphene patches shows significant intensity of the O 1s core level peak (see Supplementary Fig. S2).\\ 
The influence of oxygen is directly evident for the dominant $n_{1}$ facet. In contrast to the other two facets it generates a distinct background LEED pattern, which corresponds to an oxygen p$(2\times2)$ superlattice, $30^{\circ}$ rotated with respect to the Cu(111) reciprocal lattice (see red unit cell vectors in the LEED image), 
recently reported by Gottardi et al. in 2015~\cite{Gottardi}.\\
From the distance and alignment of the three specular spots on the LEED screen we can estimate the relative inclination angles of $n_{2}$ and $n_{3}$ with respect to $n_{1}\parallel$~Cu[111] with good accuracy. The rotation angles defined by the rotation vector [$1\overline{1}0$] amount to $(-33.2\pm 5)^{\circ}$ and $(-52.0\pm 5)^{\circ}$, respectively. At a rotation angle of $-35.3^{\circ}$ one expects the more open Cu(110) surface, which can be stabilized under the influence of adsorbates like oxygen or carbon~\cite{ViolBarbosa}. The rotation angle $(-52.0\pm 5)^{\circ}$ of the third facet's normal $n_{3}$ with respect to (111) is consistent with that of a ($22\overline{1}$) orientation, which forms an angle of $-54^{\circ}$ along the rotation direction [$1\overline{1}0$]. It is the lowest index fcc facet in the respective angle range, and can explain the homogeneous and sharp specular spot observed in LEED.\\
Our model for the nanofaceted roof-top structure in Fig.~\ref{fign1}b is further supported by $k$-PEEM images in Fig.~\ref{fign1}c performed with a 100$\mu$m-wide spot. They confirm that although graphene rotations $\varphi$ locally vary by significant relative angles, in this case $0^{\circ}$, $\pm 8^{\circ}$ and $+18^{\circ}$, the replica spots remain oriented perpendicular to the [$1\overline{1}0$] direction as expected for a homogeneous facet-induced tilting of graphene hexagons in $k$-space. In accordance with Ref.~\cite{Rasool} it suggests that graphene crystal orientations grow continuously across different facets. The $k$-PEEM pattern will be discussed in more detail below.\\

\begin{figure}
\begin{flushleft}
\center \includegraphics[width=160mm]{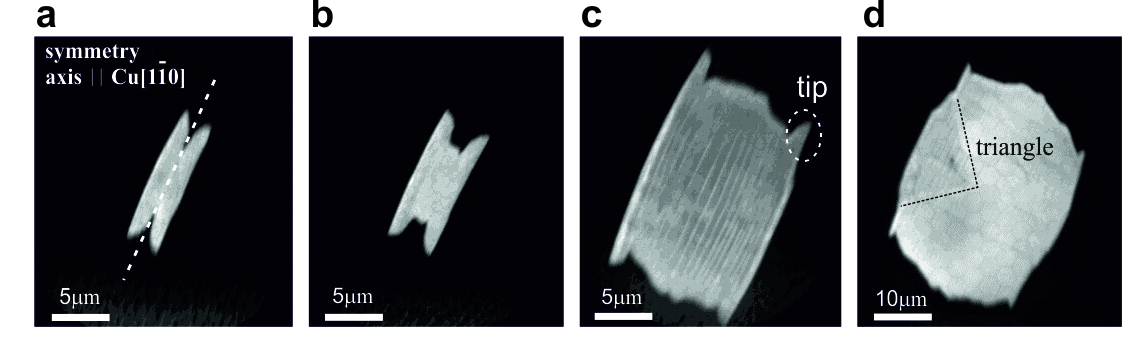}
\caption{{\bf Anisotropic growth of graphene nuclei.}
{\bf (a) - (d)} PEEM images of graphene nuclei at different stages of growth, from 2$\mu$m to 30$\mu$m width (faint hexagonal patterns are artifacts from the channel plate detector). All islands obey a twofold symmetry axis parallel to the fundamental Cu[$1\overline{1}0$] direction indicated by the dashed line in (a). Characteristic tips at the four extremities of the islands indicated in (c) are already visible at the earliest stage. Within the largest island in (d) a high-symmetry triangle is faintly visible which - as we will show in dark field contrast measurements - is due to the formation of a well-defined rotational graphene domain. 
}
\label{fign3}
\end{flushleft}
\end{figure}

{\noindent \bf Symmetry of graphene growth}.
\noindent We are interested in understanding the local influence of the faceting process on graphene growth at the earliest stage. Fig.~\ref{fign3} shows typical small graphene islands, ranging from a few $\mu$m to about 30$\,\mu$m in width. They appear bright against the dark copper oxide background due to the work function contrast in the energy-filtered PEEM imaging mode. All islands obey a two-fold mirror symmetry, and their elongation along an axis oriented parallel to the Cu[$1\overline{1}0$] direction reflects the fundamental symmetry of the faceting direction. They exhibit a characteristic tip-shaped protuberance at the four extremities, and already host the characteristic roof-top modulation structure on the $\mu$m scale. \\
Our island shapes strongly resemble those predicted recently by Meca et al. from phase-field models~\cite{Meca}, assuming markedly anisotropic carbon mobility on the metal surface. Anisotropic CVD growth usually reflects direction dependencies of e.g. chemical surface properties, or anisotropies in the morphology such as steps~\cite{Zhang, Wofford, Hayashi, Tian2012}. In our case this anisotropy is imprinted by the homogeneous vicinal Cu foil character with terraces predominantly separated by [$1\overline{1}0$]-oriented step-edges, which renders the otherwise homogeneous carbon mass transport anisotropic during the CVD process. Under the influence of CVD the vicinal structure undergoes a surface-energy driven transition to the observed complex faceted structure including the $\mu$m scale roof-top modulation. This process is facilitated by high temperatures close to the Cu melting temperature, and most likely counterbalanced by the built-up of elastic energy. \\

\begin{figure}
\begin{flushleft}
\center \includegraphics[width=14cm]{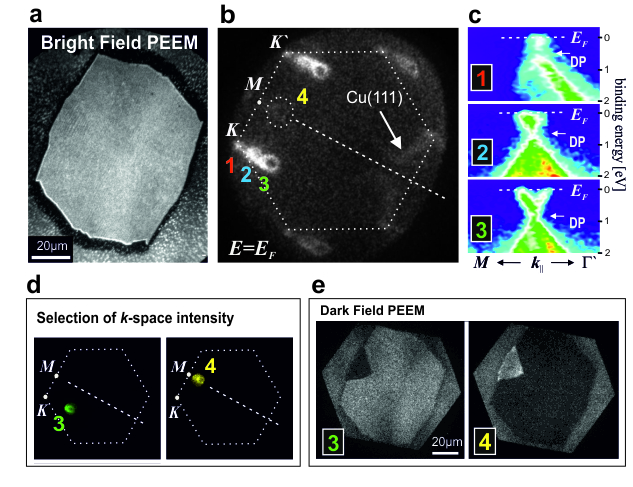}
\caption{{\bf Modulation of graphene doping levels at the nanoscale.}
Local electronic properties of a single graphene island are resolved by wave vector resolved photoemission electron microscopy. 
{\bf (a)} Bright field PEEM image of an isolated graphene island. {\bf (b)} Corresponding local $k$-space signal at $E_{\text{F}}$, showing a majority intensity at the rotational angle $\varphi=0^{\circ}$(dashed hexagon) and a faint intensity at about $\varphi=30^{\circ}$. {\bf (c)} Dispersion $E(k_{\parallel})$ of triplet replicas 1,2, and 3 ($\varphi=0^{\circ}$) at the $\text{K}$ point in the binding energy range $E_{\text{B}}=0$ (equivalent to $E_{\text{F}}$) to $E_{\text{B}}=2\,$eV. {\bf (d) and (e)} Dark field PEEM imaging at replica points 3 and 4. (d) shows aperture-selected $k$-space intensities 3 and 4 at $E_{\text{F}}$. For comparison a sketch of the $\varphi=0^{\circ}$ hexagonal graphene orientation is drawn. (e) shows according dark field contrast images revealing a highly symmetric triangular shaped $\varphi=30^{\circ}$ rotational domain embedded in the otherwise $\varphi=0^{\circ}$ host phase. 
}
\label{fign4}
\end{flushleft}
\end{figure}

{\noindent \bf Doping Modulation}. The self-organised faceting process during CVD is accompanied by a strong modulation of the electronic properties at the nanoscale. Fig.~\ref{fign4}a shows another typical larger island of about 70$\,\mu$m width. Using the spatially resolved $k$-PEEM mode of our NanoESCA instrument, the ARPES signal of the particular graphene island is captured at the Fermi level $E_{\text{F}}$, revealing the rather complex pattern shown in Fig.~\ref{fign4}b. The dominant intensity depicts the characteristic hexagon of a $\varphi=0^{\circ}$ oriented graphene domain indicated in white dashes, which exhibits the above discussed triplet replicas (denoted as 1, 2, and 3) according to the local facet planes $n_{1}$, $n_{2}$, and $n_{3}$.
A second graphene domain (denoted as 4) with minor intensity appears at $\varphi=30^{\circ}$, again locked to the symmetry of the underlying Cu(111) intensity indicated by the arrow in Fig.~\ref{fign4}b.\\
In Fig.~\ref{fign4}c the dispersion relations $E(k)$ of the three replicas at their respective $\text{K}$-points (1, 2, and 3) are shown as $(E, k)$-space cuts in the direction $\text{K}-\text{K}`$. For flat 2D graphene $k$-space information can be reduced to $k_{\parallel}$. The linear dispersion of valence and conduction bands touching at the so-called Dirac point (DP) is evident, typical of free-standing graphene with a dispersion $E(k) = \hbar v_{\text{F}}\cdot \vert k \vert$, where $v_{\text{F}}$ is the Fermi velocity. At variance with free-standing graphene however, $E_{\text{DP}}$ is not located at $E_{\text{F}}$ but shifted to higher binding energies, signifying 
electron transfer towards the graphene system, so-called $n$-doping, in accordance to previous results on graphene on homogeneous single crystalline Cu(001) and Cu(111) surfaces~\cite{Walter}. In our case substantially different doping levels $\Delta E=(E_{\text{DP}} - E_{\text{F}})$ coexist as a consequence of the three supporting facets with different interaction potentials. Starting from the hexagon 
indicated as 1 with a $n$-doping of $\Delta E = (0.4\pm0.1)\,$eV, the energy position of the DP with respect to the Fermi level $E_{\text{F}}$ shifts to larger values $\Delta E =(0.6\pm 0.1)\,$eV and $\Delta E = (0.8\pm 0.1)\,$eV for replica $\text{K}$-points 2 and 3, respectively.
The differences in doping levels within the graphene island correspond to spatial modulations of the areal carrier densities, which due to the linear dispersion close to the DP are determined by the simple relation of the density of states $D(\Delta E)=D_0\cdot \vert \Delta E \vert$, where $D_0$ is a function of the Fermi velocity $v_{\text{F}}$ (see supplementary). From the dispersion relations shown in Fig.~\ref{fign4}c we fit the Fermi velocity $v_{\text{F}}$ to values of $(0.95\pm 0.05)\times 10^6\,$m/s, giving $D_0 = 0.085$ per (eV$^2$ and graphene unit cell area) (see Supplementary Fig. S3). Integrating $D(\Delta E)$ then allows to calculate the transferred carrier densities per area in units cm$^2$ via the relation $n(\Delta E) = D_0 \cdot \Delta E^2/2$. 
Doping levels at points 1, 2, and 3 thus correspond to carrier densities of 1.6, 3.3, and 5.7 $\times 10^{13}\,$cm$^{-2}$, respectively. The large modulation $\Delta n$ by several $10^{13}\,$cm$^{-2}$ evidences the high efficiency of the nanofaceting process as a stamp for carrier modulations. \\

{\noindent \bf Dark-field characterisation}. The findings raise the question if even more detailed assignments between real space and $k$-space can be made. So far we showed ordinary bright-field PEEM images of graphene islands, where for each spot its entire $k$-space signal contributes for imaging. The electron optics of our NanoESCA instrument, however, also allows for dark-field (DF) imaging~\cite{Barrett2012}, in which a certain $k$-space intensity of interest is selected by a narrow aperture. Switching back to real space, the spatial origin of this $k$-space signal can be to traced back in the respective PEEM image. Fig.~\ref{fign4}d shows the aperture selection of the intensities at points 3 ($\varphi=0^{\circ}$) and 4 ($\varphi=30^{\circ}$) in $k$-space, while the rest of $k$-space intensity is blocked. In Fig.~\ref{fign4}e the respective DF images are presented, resolving a highly symmetric triangular $\varphi=30^{\circ}$ domain seed in a $\varphi=0^{\circ}$ graphene host structure. Measurements on many different islands on our copper foils indeed confirm that triangular seeds are oriented along the same direction perpendicular to [$1\overline{1}0$] (see Supplementary Fig.~S4). DF images at the three replica points 1, 2, and 3 show similar contrast as expected for large, $\mu$m scale continuous graphene domains on a nanoscale faceted surface.\\

{\noindent \bf \large Discussion}\\
\noindent The results of this study put forward a concept to achieve nanoscale doping modulations in chemical vapor deposited single layer graphene, exploiting surface energy driven faceting processes of supporting catalytic metals. For vicinal Cu(111) we show the self-assembly of coexisting copper ($111$), ($110$), and ($22\overline{1}$) nanofacets, which efficiently alternate graphene doping levels by several $10^{13}$~carriers per cm$^{2}$ in a periodic manner. The concept is powerful since faceting geometry and associated directional modulation of doping levels is predefined by the vicinal orientation of the catalytic metal via the conservation of total substrate symmetry~\cite{Chen2003}. Indeed, for non-vicinal Cu(111) surfaces with sixfold $C_6$ rotational symmetry the faceting effect is absent, which leads to homogeneously doped graphene sheets only. A targeted manipulation of graphene based upon the choice of vicinal symmetry can thus be envisioned, using the knowledge on various surfactant-induced faceting phenomena on different surface materials at different temperatures~\cite{Ertl}. Although in principle the equilibrium geometry of faceted surfaces are defined by the minimum of the total surface free energy, $\int{\gamma A}$, where $A$ is the area and $\gamma$ the area dependent specific surface free energy, precise predictions are often hampered by the fact that systems not always reach their thermodynamic equilibrium due to kinetic barriers in the faceting process. Nevertheless, due to the vast available parameter space in surface science our concept is potent and can be generalized to other chemical vapor deposited 2D systems in the focus of present research, such as semiconducting transition metal dichalcogenides, e.g. MoS$_2$, insulating hexagonal boron nitride (h-BN) monolayers. \\
For graphene on vicinal Cu(111) studied in this work, we propose that the complex bottom-up faceting process is the result of anisotropic mobilities of carbon and copper atoms during high-temperature CVD close to the copper melting temperature. Mobilities and thus mass transport are asymmetric along and perpendicular to the vector [$1\overline{1}0$] characterising the vicinal surface orientation. This leads to strictly aligned graphene island nuclei with twofold $C_2$ rotational symmetry similar to those predicted recently by theory on CVD kinetics under anisotropic growth conditions~\cite{Meca}. Of vital importance for the understanding of the evident correlation between spatial and electronic band structure symmetries is the local wave-vector-resolved photoemission microscopy ($k$-PEEM) technique, which enables to detect both real space and $k$-space signals of a particular graphene island of a micrometer scale. Boundaries between rotational graphene domains resolved in the dark-field mode of $k$-PEEM involve abrupt changes of the faceting morphology (see Supplementary Fig.~S4, (b)-(d)), which underlines the intimate feedback between graphene and copper during the surface energy driven faceting processes. \\

{\noindent \bf \large Methods}\\

\begin{footnotesize}

\noindent {\bf Sample preparation}\\
For the CVD growth of graphene islands we use $25\,\mu$m thick copper foils ($99.8\,\%$ purity, Alpha
Aesar, reference 13382). A protective oxide is stripped by electrolysis in a copper sulfate solution before graphene grains are grown by CVD using the pulsed method as described in Ref.~\cite{Han}. 50~mm wide pieces of Cu foil are loaded into the CVD reactor then heated up to $1020\,^{\circ}$C under a $50\,$sccm of argon at 3~mbar. Growth conditions are obtained by a series of 72 pulses of methane 4~sccm during $12\,$sec separated by 50~sec long idling times. Argon and hydrogen input are kept constant during growth and respectively equal to 50 and $1000\,$sccm. The pressure is $3\,$mbar during the full process.\\
Prior to surface sensitive UHV analysis techniques XPS, $k$-PEEM, LEED, and STM the samples were annealed under UHV conditions for 30~min at temperatures of $400\,^{\circ}$C to ensure cleanliness. If needed, graphene was removed by mild Ar$^+$-ion sputtering at room temperature ($10^{-6}\,$mbar argon pressure, cathode voltage 1~kV, 40~min) followed by $400^{\circ}$C annealing under UHV.\\

\noindent {\bf Measurement techniques}\\
XPS and $k$-PEEM measurements were done using an {\it Omicron NanoESCA} instrument with laboratory light sources. The photoemission spectrometer is based on a PEEM column and a imaging double hemispherical energy filter~\cite{Kromk}. A transfer lens in the electron optics switches between real space and angular resolved $k$-PEEM mode, which allows to detect classical x-ray photoemission spectra with monochromatized Al $K_\alpha$ radiation, as well as an energy dependent mapping of the Brillouin zone using a helium discharge lamp at $h\nu=21.2\,$eV with a resolution of 
$\Delta E \approx 0.2\,$eV. In the $k$-PEEM mode the Fermi edge $E_{\text{F}}$ is derived from the kinetic energy at which the k-PEEM intensity is cut off. We define this energy as zero binding energy and expect the error of this Fermi level estimation to be $\pm 0.05\,$eV.\\ 
For dark field measurements, apertures sizes $150\,\mu$m were positioned to select intensities of interest in $k$-space. Thereafter, dark field PEEM images were taken in the respective telescopic mode. In order to minimize parasitic signals, the iris aperture was narrowed down to encompass the graphene island of choice. See also Ref.~\cite{Barrett2012} for further details.\\
Electron backscatter diffraction (EBSD) data was aquired with an {\it EDAX} camera and software on a {\it Tescan FERA 3} instrument. Detailed technical information can be found in Ref.~\cite{Wright2015}. For atomic force microscopy (AFM) a {\it Bruker Dimension Icon} was employed. Both EBSD and AFM were performed under ambient conditions.
Scanning tunneling microscopy (STM) was done at room temperature under UHV conditions ($p \approx 5\times 10^{-11}\,$mbar) using an {\it Omicron VT-STM}.\\

\end{footnotesize}

{\noindent \bf \large Acknowledgements}\\
\noindent We thank Konrad Winkler from Omicron for helpful discussions. The authors acknowledge financial support from the Ministry of Education, Youth and Sport of the Czech Republic (Research infrastructures 
SAFMAT LM2011029 and LO1409), and the EU Graphene Flagship. JH acknowledges the Purkyn\v{e} Fellowship program of the Academy of Sciences of the Czech Republic.\\

{\noindent \bf \large Author Contributions}\\
\noindent D. K. prepared the samples. M. V. measured $k$-PEEM and dark field PEEM. L. F. and M. K. provided scanning force and tunneling microscopy images, respectively. J. K. contributed with EBSD data. J. H. prepared the main manuscript text and figures together with J. C.. J. L. and V. B. reviewed the manuscript.\\

{\noindent \bf \large Additional Information}\\
\noindent {\bf Competing financial interests}: The authors declare no competing financial interests.\\

\clearpage

\clearpage

\end{document}